\documentclass[]{spie}  %>>> use for US letter paper
\usepackage{amsmath,amsfonts,amssymb}
\usepackage{graphicx}
\usepackage[colorlinks=true, allcolors=blue]{hyperref}
\usepackage{siunitx}
\usepackage{todonotes}
\usepackage{cleveref}
\usepackage{subcaption}

\title{Numerical field optimization for enhanced efficiency in time-reversible gradient computation of open-source GPU-accelerated FDTD simulations}

\author[a]{Yannik Mahlau}
\author[a]{Lukas Berg}
\author[a]{Bodo Rosenhahn}
\affil[a]{Institute of Information Processing, Leibniz University Hannover, Appelstraße 9a, 30167 Hannover, Germany}

\authorinfo{Further author information: (Send correspondence to Yannik Mahlau)\\Yannik Mahlau: E-mail: mahlau@tnt.uni-hannover.de, Telephone: +49 511 762-5316}

\begin{document} 
\maketitle

\begin{abstract}
Finite-difference time-domain (FDTD) simulations often involve physical quantities spanning multiple orders of magnitude, such as the speed of light or electromagnetic field amplitudes.
The standard practice for maintaining numerical accuracy in many FDTD implementations is to use 32-bit or 64-bit floating-point values to represent the electric and magnetic fields.
However, this approach is not always optimal when recording field values, particularly during time-reversible gradient computation where electric and magnetic field values need to be saved at the boundary of the simulation domain.
Since this memory bottleneck is often the limiting factor in time-reversible inverse design for nanophotonics, we present two field optimizations for enhancing memory efficiency in FDTD simulations.
Using a smaller bit-width representation of field values as well as interpolation, we achieve similar accuracy at lower memory cost.
This approach is particularly beneficial for GPU-accelerated computing, where reduced-precision data types are increasingly preferred due to their computational efficiency and prevalence in machine learning frameworks.
We integrate our approach into FDTDX, an open-source, differentiable FDTD solver that natively supports time-reversible gradient computation.
Our approach is especially important for future developments towards large-scale open-source simulations, which are critical for advancing computational nanophotonic applications.
\end{abstract}


\section{Introduction}
Inverse design in nanophotonics \cite{molesky2018inverse} is a growing field since it enables the automated design of high-performance components without manual intervention.
It encompasses both free-form topology optimization \cite{gedeon2023time, schubert2022inverse, mahlau_rl} and shape optimization \cite{schneider2019benchmarking, mahlau2026bonni} with fixed shape parameterizations of photonic nanostructures.
In inverse design, gradient computation is typically performed using the adjoint method \cite{luce2024merging, pontryagin2018mathematical}, which only requires one additional simulation to calculate the gradient.

Recently, time-reversible gradient computation \cite{tang2023time, schubert2025quantized} has emerged as a powerful technique, as it enables the gradient calculation of arbitrary time-dependent objective functions.
This opens the possibility to apply inverse design to many applications such as pulse shaping \cite{geromel2023compact, shi2026ultrafast}, supercontinuum generation \cite{montesinos2020chip, singh2018octave} or spatio-temporal metamaterials \cite{camacho2020achieving}.
Similar to the adjoint method, this approach also requires only a single additional simulation.
It begins at the final time step of the forward simulation and propagates backward through time \cite{blondel2024elements}.
This allows for memory-efficient gradient computation through the time domain, because it eliminates the need to save electric and magnetic fields at every time step of the entire simulation volume.
However, since absorbing boundaries such as Perfectly Matched Layers (PML) \cite{roden2000convolution} are non-invertible in time, field values for these boundaries need to be saved during the forward simulation \cite{mulder2018higher}.
For long simulations, storing this field data is highly memory-intensive and becomes the main bottleneck for scaling simulation size.
To alleviate this issue, we introduce two straightforward techniques to improve memory efficiency.
By reducing the field data bit-width and subsampling every $k$ time steps, we drastically reduce the memory requirements of gradient computation while keeping reasonable accuracy.
This is particularly beneficial for GPU-accelerated simulations \cite{mahlau_flexible, tidy3d}, where memory is expensive and reduced-precision data types are often preferred due to their prevalence in machine learning frameworks \cite{jax2018github}.
To summarize, our main contributions are
\begin{itemize}
    \item We present two compression techniques for reducing memory requirements of time-domain gradient computation: temporal subsampling of the fields and reducing their bit-width representation.
    \item Using these techniques, we are able to show a reduction in memory requirements of \textbf{64x} without result degradation.
    Because these savings scale with simulation resolution, optimizations at finer grid sizes will yield even greater benefits.
    \item To facilitate adoption of our work, we integrate the compression techniques into the open-source solver FDTDX \cite{fdtdx_joss}.
\end{itemize}

\section{Methods}

\begin{figure}[!t]
    \centering % Centers the subfigures on the page
    \begin{subfigure}[b]{0.48\textwidth}
        \includegraphics[width=\linewidth]{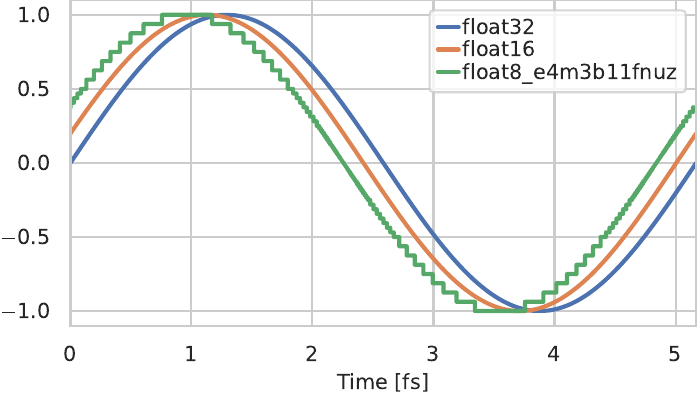}
        \caption{Data Type Conversion}
        \label{fig:sine_dtype}
    \end{subfigure}
    \hfill
    \begin{subfigure}[b]{0.48\textwidth}
        \includegraphics[width=\linewidth]{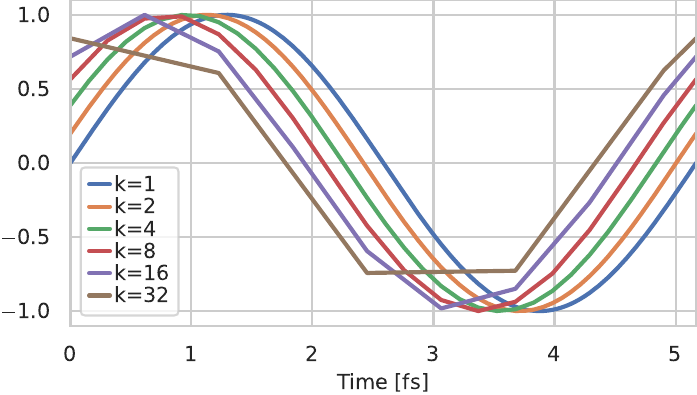}
        \caption{Linear Interpolation}
        \label{fig:sine_interpolation}
    \end{subfigure}
    \vspace{0.1cm}
    \caption{Compression of an original sine wave (blue) in 32 bit precision (a) and 136 time steps per period (b).
    The original wave is compressed to a smaller data type (a) or subsampled at every k time steps (b).
    The phase offset between different curves in the plots is added for visualization purposes and has no connection to the compression.
    }
    \label{fig:sine}
\end{figure}

For time-reversible gradient computation, the field values at the interface slices between the simulation volume and PML must be recorded in the forward simulation.
In the time-reversed simulation, these field values are injected back at the respective interface slices.
Effectively, the boundary interfaces act as sources during the time reversed simulation.
The standard bit-widths for numerical simulations are \texttt{float32} or \texttt{float64} \cite{ieee1985ieee, muller2018handbook}, also called single and double precision, respectively.
Using these large bit-widths is necessary to ensure accuracy \cite{lesina2015convergence, taflove2005computational}.
However, it may not be necessary to use a large bit-width for storing field values if they are not directly used in any computation.
To reduce memory overhead, we propose converting the recorded field values to smaller bit-widths such as \texttt{float16} (half precision) or \texttt{float8\_e4m3b11fnuz} \cite{kalamkar2019study}.
We convert the stored values back to the larger bit-widths of either single or double precision during time-reversed simulation.
Therefore, there is no direct computation performed on smaller bit-widths.
Instead, the only loss in accuracy stems from the conversion error when converting values of large to smaller bit-width.

Furthermore, depending on the spatial resolution and source frequency, saving field values at every time step is often unnecessary.
When running a three-dimensional FDTD simulation with a uniform spatial resolution of $\Delta x$, the Courant-Friedrichs-Levy (CFL) stability condition \cite{courant1928partiellen} dictates a maximum time step of
\begin{align}
    \Delta t = \frac{C\, \Delta x}{c\, \sqrt{3}},
\end{align}
where $C = 0.99$ is a typical Courant factor and $c = 299792458 \,\frac{\text{m}}{\text{s}}$ is the speed of light in vacuum.
For a resolution of $\Delta x = 20$ nm, this yields a time step of $\Delta t = 3.81 \times 10^{-17}$ seconds.
A typical infrared source of 1550 nm has a period of $5.17 \times 10^{-15}$ seconds, which corresponds to approximately 136 time steps.
To accurately reconstruct the electric and magnetic fields of such a wave, recording every time step is redundant.
Instead, subsampling at every $k$-th time step and applying linear interpolation provides sufficient accuracy.
To illustrate the effect of reduced bit-width and temporal subsampling, we represent a sine wave using both techniques in \cref{fig:sine}.

\begin{figure}[!t]
    \centering
    \includegraphics[width=0.8\linewidth]{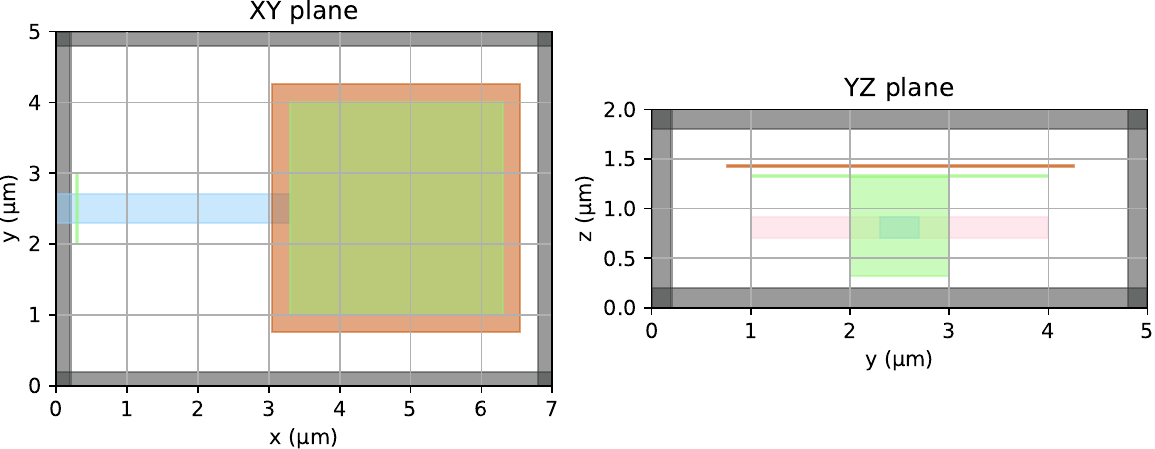}
    \vspace{0.1cm}
    \caption{Simulation setup for the inverse design of a grating coupler.
    A source (orange) injects light into the simulation towards the design region (pink).
    The design is optimized such that light is redirected into the output waveguide (blue).
    The ratio between input and output poynting flux is measured through two detectors (green).
    The simulation volume is surrounded by PML boundary objects (grey). 
    }
    \label{fig:setup}
\end{figure}

\section{Results and Discussion}
We evaluate our methods on the topology optimization of a silicon coupler \cite{huang2025compact}, designed to direct free-space light into a waveguide.
Because this application lacks a time-dependent objective function, it could alternatively be optimized using standard adjoint gradient calculations.
Nevertheless, we select it as a straightforward benchmark to validate our approach.
The coupling device measures $1.6$ µm along both the x- and y-axes, with a height of 220 nm.
The connecting waveguide shares this 220 nm height and has a width of 400 nm.
A source positioned directly above the device illuminates it with a normally incident plane wave.
To calculate transmission efficiency, we measure the Poynting flux both above the device and within the output waveguide.
The device itself is parameterized by continuous variables in the range $[0, 1]$ mapped to the materials silicon and silicon dioxide.
This mapping relies on a Gaussian filter with a 60 nm standard deviation, followed by a subpixel-smoothed projection.
This smoothed projection enables gradient-based optimization even with a projection parameter of $\beta = \infty$, since it combines level-set and density-based projection methods \cite{hammond2025unifying}.
\Cref{fig:setup} illustrates the complete simulation setup.

\begin{figure}[!t]
    \centering
    \includegraphics[width=0.58\linewidth]{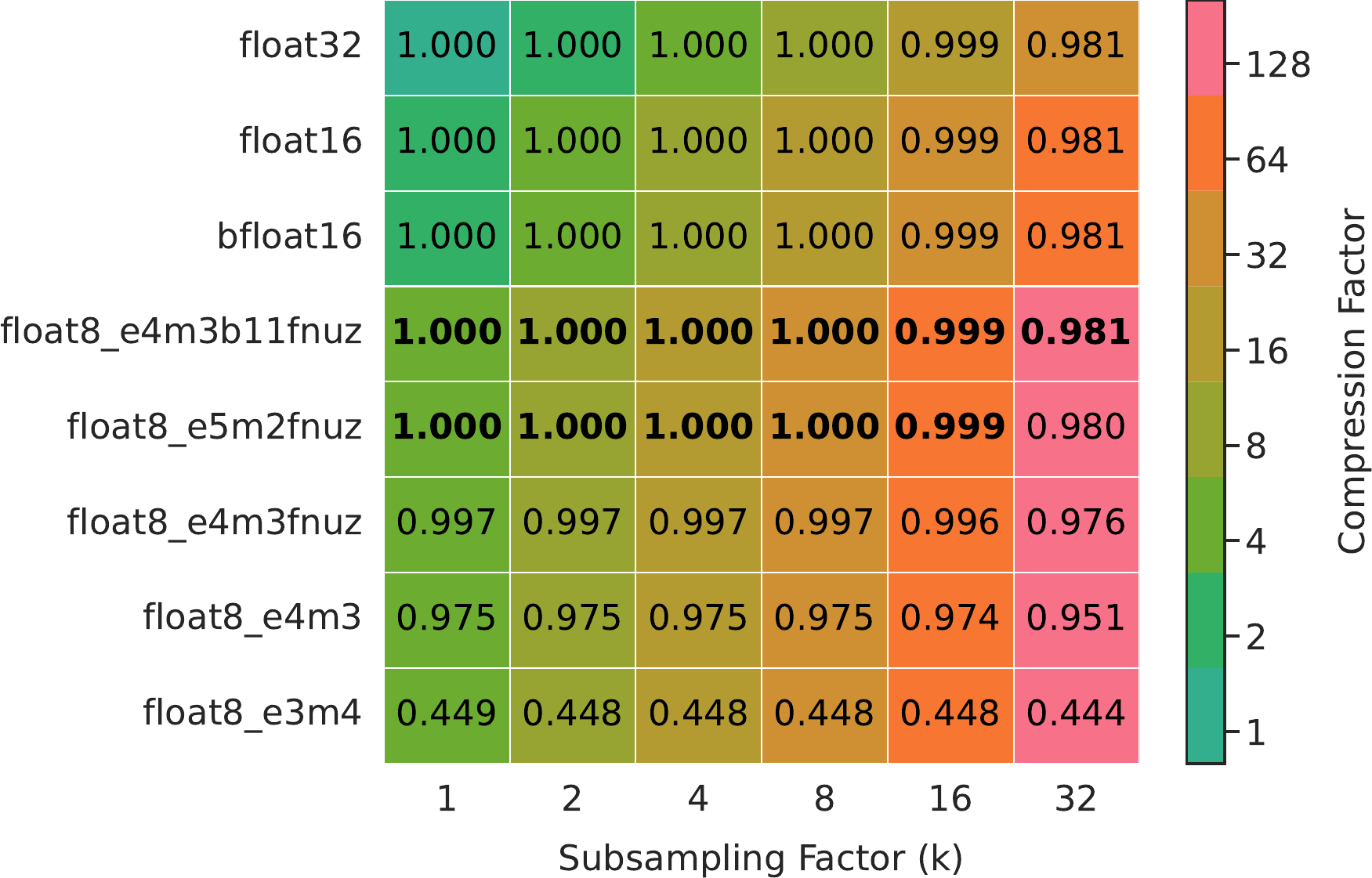}
    \vspace{0.1cm}
    \caption{Mean cosine similarity of gradient computation using different data types and subsampling factors.
    Using \texttt{float32} with recording every time step ($k = 1$) as a baseline, the compression factor is the memory saved through lower bit-width multiplied with $k$.
    The mean accuracy is calculated over 10 gradient calculations.
    The best results per subsampling factor are marked in bold.
    }
    \label{fig:compression}
\end{figure}

We first evaluate the similarity between baseline gradients calculated at full precision without subsampling and those computed using various subsampling factors $k$ and reduced-precision data types.
We test subsampling factors ranging from 1 to 32.
For data types, we test full precision \texttt{float32} as well as half precision \texttt{float16} and \texttt{bfloat16} \cite{kalamkar2019study, burgess2019bfloat16}.
The \texttt{bfloat16} data type has a larger range of representable values than \texttt{float16} at the cost of reduced precision for small numbers.
Additionally, we evaluate different \texttt{float8} data types \cite{micikevicius2022fp8}, namely \texttt{float8\_e4m3b11fnuz}, \texttt{float8\_e5m2fnuz}, \texttt{float8\_e4m3fnuz}, \texttt{float8\_e4m3}, and \texttt{float8\_e3m4}.
These data types are distinguished by the number of bits allocated for exponent, mantissa as well as the size of the implicit bias used for exponent representation.
The suffix \texttt{f} signals that only finite values are represented.
Furthermore, \texttt{n} denotes support for Not-a-Number (NaN) values, while \texttt{uz} specifies an unsigned zero representation.
To assess gradient quality, we uniformly sample design parameters and calculate the corresponding gradients using different sampling factors and data types.
Since modern optimizers such as Adam \cite{kingma2014adam} are robust to varying gradient magnitudes, we are mainly interested in the gradient direction.
Therefore, we measure the cosine similarity 
\begin{align}
    S(x, y) = \frac{x \cdot y}{||x|| \cdot ||y||}
\end{align}
 between the baseline gradient (\texttt{float32} and $k=1$) and gradients computed with reduced precision and subsampled field recordings.

In \cref{fig:compression}, the results are visualized.
We observe no degradation in gradient similarity for subsampling factors up to $k=8$ with data type \texttt{float32}.
Similarity drops slightly at $k=16$, and decreases more significantly at $k=32$.
When using \texttt{float16}, \texttt{bfloat16}, and \texttt{float8\_e4m3b11fnuz}, the gradient similarity remains indistinguishable from the \texttt{float32} baseline.
The \texttt{float8\_e5m2fnuz} format performs similarly, though minor deviations emerge at $k=32$.
The remaining data types exhibit substantial degradation in gradient similarity across all subsampling factors.
This experiment demonstrates that selecting the appropriate 8-bit representation is crucial for preserving numerical accuracy.
Furthermore, it indicates that inverse design optimizations can achieve a 64-fold compression factor using \texttt{float8\_e4m3b11fnuz} and $k=16$ without compromising gradient quality.

\begin{figure}[!t]
    \centering % Centers the subfigures on the page
    \begin{subfigure}[b]{0.55\textwidth}
        \includegraphics[width=\linewidth]{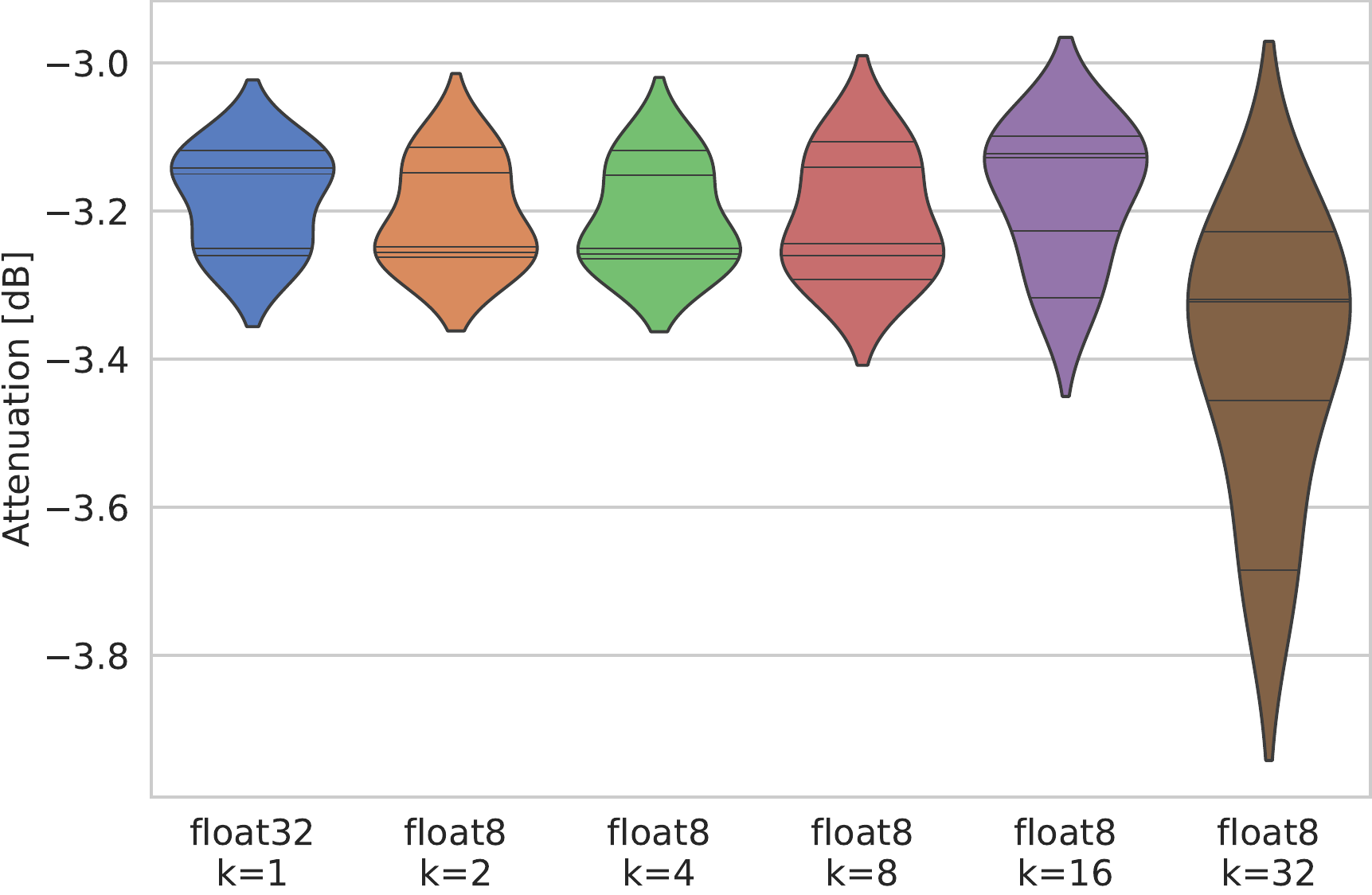}
        \caption{Optimization Results}
        \label{fig:optim_results}
    \end{subfigure}
    \hfill
    \begin{subfigure}[b]{0.35\textwidth}
        \raisebox{1cm}{\includegraphics[width=\linewidth]{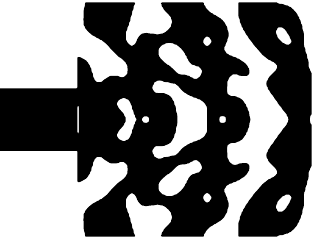}}
        \caption{Optimized Design}
        \label{fig:optim_design}
    \end{subfigure}
    \vspace{0.1cm}
    \caption{Results of inverse design optimizations of a grating coupler.
    In (a), the transmission attenuation of designs optimized using data types \texttt{float32} or \texttt{float8\_e4m3b11fnuz} with different values of $k$ are shown.
    The five thin horizontal lines indicate the results of five runs started at random parameters.
    The best result obtained by using \texttt{float8\_e4m3b11fnuz} and $k=16$ is visualized in (b).
    }
    \label{fig:optim}
\end{figure}

However, the practical impact of these gradient deviations on the final inverse design outcome remains unclear.
To investigate this, we compare the baseline optimization (\texttt{float32}, $k=1$) against memory-efficient configurations utilizing \texttt{float8\_e4m3b11fnuz} and various values of $k$.
We execute the optimization over 200 gradient steps, applying a projection value of $\beta = 1$ for the initial 40 iterations and $\beta = \infty$ thereafter.
For gradient updates, we employ the Adam optimizer \cite{kingma2014adam} with Nesterov momentum \cite{dozat2016incorporating} and a linear warmup learning rate schedule \cite{goyal2017accurate}.
To reduce statistical dependence on the starting conditions, we repeat the optimizations across five randomly sampled sets of initial design parameters.
\Cref{fig:optim} presents these optimization results.
While the initial designs exhibit a transmission of roughly -25 dB, all optimized designs reach a transmission between -3 dB and -4 dB.
We observe only minor deviations between the baseline optimization and the reduced-precision runs with $k \leq 16$.
Performance deviations become more pronounced only at $k=32$.
However, these variances remain small in comparison to the large improvements over the initial random designs.
This indicates that the Adam optimizer is capable of optimizing designs with stochastic gradient errors.
This resilience is expected, given that Adam was originally developed to handle stochastic batch gradients in machine learning applications \cite{kingma2014adam}.
Interestingly, the single best result is achieved using \texttt{float8\_e4m3b11fnuz} with $k=16$, slightly outperforming the memory-intensive baseline.
This suggests that small gradient perturbations could aid the optimization process, a regularization phenomenon well-documented in the field of machine learning \cite{jin2017escape, neelakantan2015adding}.
However, further investigation is required to draw conclusive claims for this specific application.

\section{Conclusion}
We demonstrated two compression techniques to improve memory efficiency in the time-reversible gradient computation of FDTD simulations.
By employing reduced-precision data types like \texttt{float8\_e4m3b11fnuz} and linear interpolation between recorded field values, we reduce the memory footprint of gradient computation significantly while maintaining high gradient accuracy.
Furthermore, evaluations using the Adam optimizer reveal that the minor gradient errors introduced by reduced precision and temporal subsampling have a negligible impact on final design quality.

For future work, we plan to develop advanced field compression techniques inspired by standard image compression algorithms, such as JPEG \cite{wallace1991jpeg}.
Ultimately, this enhanced efficiency will enable the application of inverse design to much larger and more complex systems.
By removing this computational bottleneck, researchers will be equipped to tackle intricate, multi-objective design problems that were previously intractable in nanophotonics.

\section*{Acknowledgments}
This work was supported by the Federal Ministry of Education and Research (BMBF), Germany, under the AI service center KISSKI (grant no. 01IS22093C), the Deutsche Forschungsgemeinschaft (DFG) under Germany’s Excellence Strategy within the Clusters of Excellence PhoenixD (EXC2122) and Quantum Frontiers-2 (EXC2123), the European Union  under grant agreement no. 101136006 – XTREME.
Additionally, this work was funded by the Deutsche Forschungsgemeinschaft (DFG, German Research Foundation) – 517733257.

% References
\bibliography{report} % bibliography data in report.bib
\bibliographystyle{spiebib} % makes bibtex use spiebib.bst

\end{document}